\documentclass{iopjournal}
\usepackage{amsmath}
\usepackage{tabularx}
\usepackage{multirow}
\usepackage{array}
\usepackage{booktabs}
\usepackage{graphicx}
\usepackage{dcolumn}
\usepackage{bm}
\usepackage{makecell}
\begin{document}
\pagestyle{empty}
\title{A Novel Scheme for Inter-Satellite Integrated Laser Ranging and Communication in Space-Based GW Detection}
\author{Zihao Shao$^{1,2}$, Siyuan Xie$^2$, Kai Luo$^{2,*}$}
\affil{$^1$National Gravitation Laboratory, MOE Key Laboratory of Fundamental Physical Quantities Measurement, and School of Physics, Huazhong University of Science and Technology, Wuhan 430074, China}
\affil{$^2$School of Electronics and Communication Engineering, Sun Yat‑sen University, Shenzhen 518107, China}
\affil{$^*$Author to whom any correspondence should be addressed.}
\par
\email{kluo@mail.sysu.edu.cn}
\begin{abstract}
Adopting pseudo‑random noise (PRN) codes for inter‑satellite ranging, the resulting data‑encoded PRN signal induces excessive laser phase measurement noise under the currently adopted integrated scheme in space‑based gravitational wave (GW) detection systems. To suppress the noise, a scheme of inter‑satellite integrated laser link is presented in this Letter, in which the data‑encoded PRN signal for inter‑satellite absolute distance measurement and communication is phase‑modulated onto the ultra‑stable oscillator (USO) clock signal before phase modulation onto the laser carrier. Theoretical analysis and simulation results show that in the proposed scheme the noise is significantly suppressed compared with the currently‑adopted one, while the noise stemming from the data‑encoded PRN signal on clock noise transfer is limited within the requirement.
\end{abstract}
\vspace{6pt}
\section{Introduction}
\begin{figure}[ht!]
 \centering
        \includegraphics[width=8.5cm,keepaspectratio]{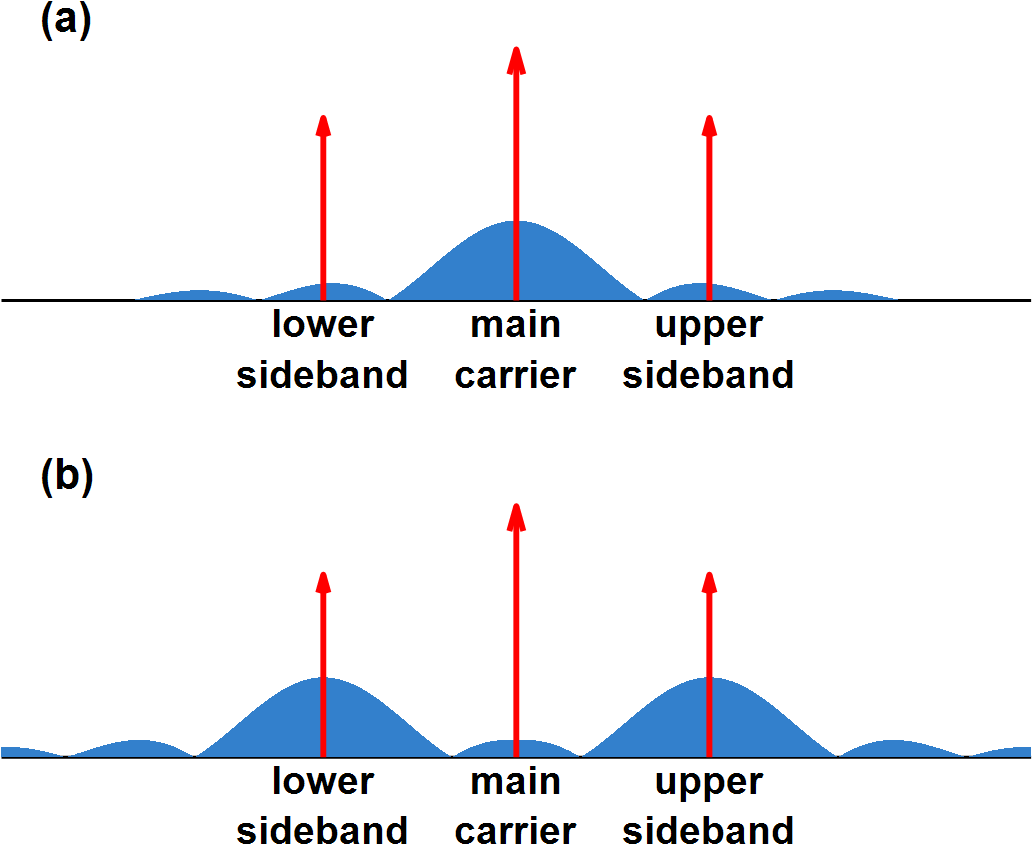}
 \caption{Conceptual schematic diagrams of (a) the currently‑adopted carrier phase modulation based integrated scheme and (b) the proposed sideband modulation based scheme. In both panels, the blue regions represent the spectrum of the data‑encoded PRN signal, while the red arrows denote the main carrier and the clock sidebands of the interferometric signal.}
\label{fig1}
\end{figure}

Space‑based gravitational wave (GW) detection is able to access the low‑frequency band below\,1\,Hz to achieve unique observations of massive black hole mergers, galactic binaries, and primordial signals from the early Universe~\cite{Danzmann_1996, Luo_2015,amaroseoane2017laser}, compared with ground‑based detection that is sensitive to the high frequency band above\,10\,Hz ~\cite{Harms_2015, PhysRevD.95.103012}. During the data post‑processing of space‑based GW detection, the time‑delay interferometry (TDI) technique is used to suppress laser frequency noise~\cite{Armstrong1999TDI, Estabrook2000LISAdata, Tinto2021TDI}, with the absolute inter‑satellite distances among the three spacecraft required as input~\cite{Muratore2020LaserNoise, Wang_2024}. For this purpose, inter‑satellite ranging is carried out using a specified ranging signal with an appropriate ambiguity range and $1\,\mathrm{m}$ ranging accuracy~\cite{2011OExpr..1915937J, Sutton2010, Heinzel2011}. Meanwhile, the ultra‑stable oscillator (USO) clock signals need to be exchanged among satellites to remove in‑band clock jitter~\cite{Tinto2002LISATDI, Tinto2018ClockNoise, Hartwig2022}. Furthermore, inter‑satellite communication is required to support cooperative data processing across the satellite constellation~\cite{Esteban2009OpticalLink, Liang2025LaserSim}. In light of these aspects, space‑based GW detection systems rely on inter‑satellite absolute ranging, communication, and clock noise transfer as auxiliary functions of the laser link. Given constrained onboard payload and the requirement of simultaneous phase, ranging, and differential clock noise for TDI, laser phase measurement and these auxiliary functions must be integrated to share the same laser link~\cite{Sutton2010, Heinzel2011, Otto_2012, Yamamoto2024LISA_ranging}. Extensive research has been devoted to integrated schemes combining these functions that meet the extreme precision requirements of space‑based GW detectors.
Single‑frequency ranging tones can be employed for laser ranging~\cite{Poujouly_2002, He:20}. For space‑based GW detection missions, they were once phase‑modulated onto the USO clock signal along with the communication signals and the resulting signal was subsequently phase‑modulated onto the laser carrier so that the impact of auxiliary functions on laser phase measurement of the detection system was minimized~\cite{Pollack_2006}. Later, ranging tones were replaced with pseudo‑random noise (PRN) codes for larger unambiguity range and superior anti‑interference performance~\cite{Nie_2020, Zhi_2025}, and the codes were combined with communication signals using direct‑sequence spread spectrum (DSSS) modulation to produce the data‑encoded PRN signal. The integrated scheme currently adopted was developed by LISA~\cite{Heinzel2011}. It modulates the USO clock signal and data‑encoded PRN signal onto the laser carrier through phase modulation, thereby realizing the integration of the auxiliary functions required for the detection system. To achieve mission‑specified accuracy for the detection of targeted GWs, laser phase measurement noise must be constrained to the picometer level~\cite{Wissel_2023, Xu_2025}. Consequently, for such an integrated scheme, particular attention is paid to laser phase measurement noise originating from auxiliary functional components within the millihertz‑to‑hertz band of space‑based GW detection. Among these noise sources, the data‑encoded PRN signal employed for inter‑satellite ranging and communication ranks among the most dominant contributors~\cite{shaddock2006phasemeter, Colpi2024, Li2025}. Numerous studies have analyzed the noise induced by the signal in laser phase measurement under this scheme~\cite{Euringer2024, Heinzel2011, 2011OExpr..1915937J, Sutton2010}. It has been demonstrated that the signal component of the data‑encoded PRN signal modulated onto the laser carrier exceeds the allowable system noise budget, which further limits the laser phase measurement precision of GW detection systems. Owing to timing jitter of the USOs on the two spacecraft within the same link, the USO‑related contributions in the data‑encoded PRN signals cannot be completely eliminated merely by laser carrier phase‑locking and TDI techniques. Whether these residual contributions after TDI processing will exceed the detection system requirements remains to be determined in future studies. To alleviate this noise, existing efforts have been focused on PRN code design. Manchester codes are proposed by LISA to reduce the low‑frequency components of the data‑encoded PRN signal, thus suppressing the noise~\cite{Sutton2013OpticalRanging, Esteban2012LaserPhD, 2007PhDT.......260W}. However, two relatively large sidelobes in the autocorrelation function of Manchester codes may raise the probability of false acquisition during code acquisition, thereby severely degrading the ranging accuracy of the detection system. To further improve the performance of PRN codes, LISA has also proposed PRN structures with reduced millihertz‑band energy leakage such as bit‑balanced codes (BBC), yet this performance enhancement is achieved at the cost of degraded autocorrelation properties~\cite{Yamamoto2023ClockSync}.
In this Letter, departing from the complicated code design for the currently‑adopted carrier phase modulation based integrated scheme, a sideband modulation based scheme of inter‑satellite integrated laser link is proposed to suppress the noise induced by the data‑encoded PRN signal. In this scheme, the data‑encoded PRN signal is phase‑modulated onto the USO clock signal to reduce its component around the main carrier, thereby yielding superior overall integration performance over the currently‑adopted one. To clearly illustrate the principle and structural differences between the currently‑adopted and proposed schemes, conceptual schematics of both schemes are presented in Fig.~\ref{fig1}.
\section{Signal Modeling and Scheme Analysis}
\begin{figure}[ht!]
 \centering
        \includegraphics[width=8.5cm,keepaspectratio]{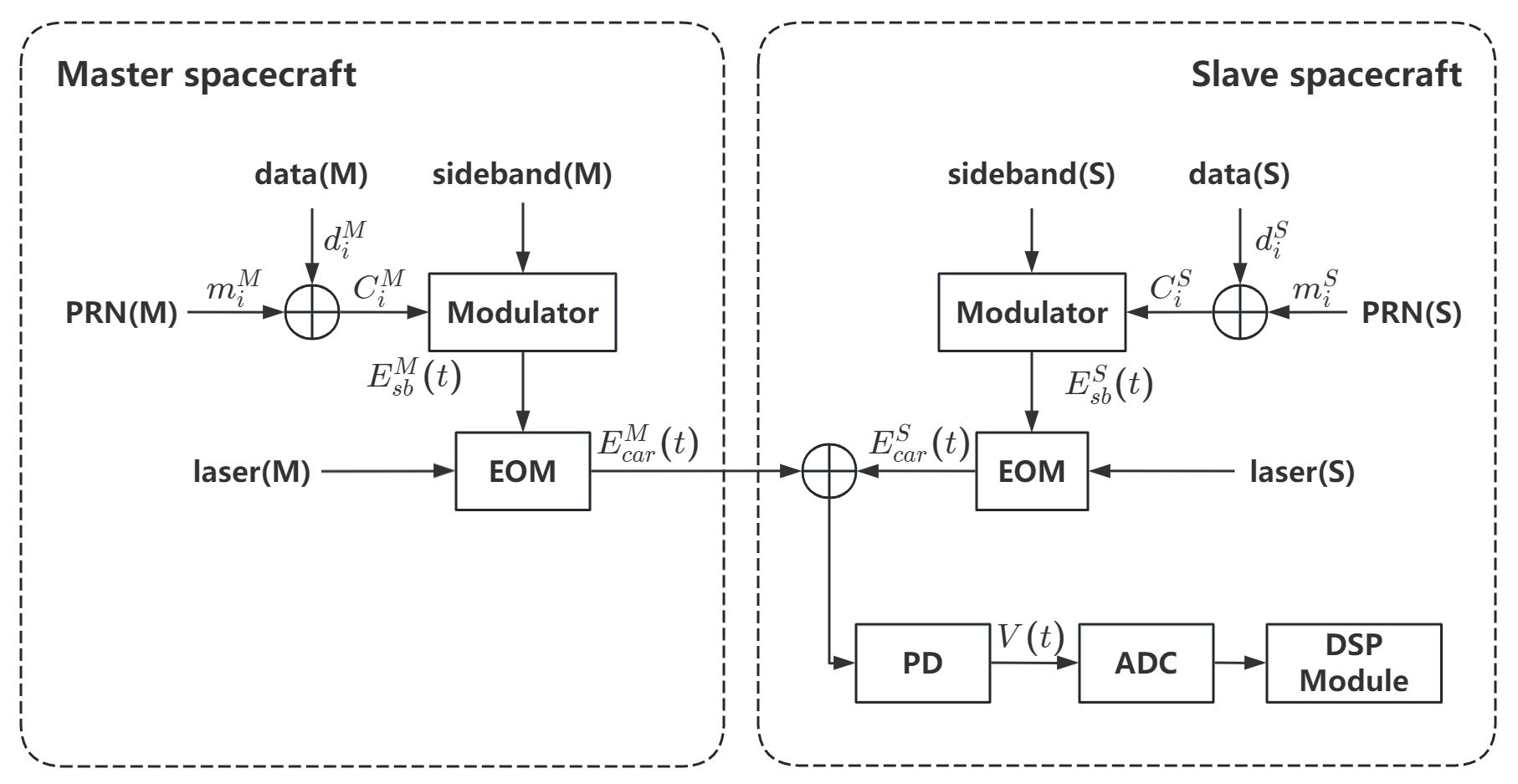}
 \caption{Generation flowchart of the transmit integrated signal in the sideband modulation based scheme of inter‑satellite integrated laser link. }
\label{fig2}
\end{figure}

A more practical design of the proposed scheme of inter‑satellite integrated laser link is shown in Fig.~\ref{fig2}.
Considering the inter‑satellite link from the master spacecraft to the slave spacecraft, the data‑encoded PRN signal is phase‑modulated onto the USO clock signal at the depth $m_{\mathrm{prn}}$~\cite{Sutton2010, Esteban2012LaserPhD}, rather than the laser carrier, to form the modulated sideband signal as Eq.~\eqref{eq:1}. For simplicity, the modulated sideband signal is normalized to unit amplitude~\cite{Glover_2009}.
\begin{equation}\label{eq:1}
\begin{split}
E_{\mathrm{sb}}^M(t)=\sin \left[2 \pi f_{\mathrm{sb}}^M t+m_{\mathrm{prn}} \cdot \phi _{\mathrm{seq}}^{M}(t) \right]
\end{split}
\end{equation}
where $f_{\mathrm{sb}}^M$ is the USO clock frequency of the master spacecraft. Let the chip duration be $T$ and the pulse shape be $p(t)$, then the data‑encoded PRN signal in Eq.~\eqref{eq:1} can be expressed as $\phi _{\mathrm{seq}}^{M}(t) = \sum\limits_{i=1}^{L} C_i^M \cdot p(t-i T)$, where $\{C_i^M\}$ is the data‑encoded PRN sequence obtained by exclusive‑oring (XORing) the PRN sequence of length $L$ and the communication data with a spreading factor $N$, following $C_{(k-1) N+n}^M=m_{(k-1) N+n}^M \oplus d_k^M, k=1,2 \ldots \frac{L}{N}, n=1,2 \ldots N$.
Then, the modulated sideband signal is further phase‑modulated onto the laser carrier at the depth $m_{\mathrm{sb}}$, producing the transmit integrated signal with all auxiliary functions. The transmit integrated signal $E_{\mathrm{car}}^M(t)$ from the master spacecraft can be expanded via the Jacobi‑Anger expansion~\cite{2020JCoPh.41609489L} as
\begin{equation}\label{eq:2}
\begin{split}
E_{\mathrm{car}}^M(t)
&=\sqrt{P_M} e^{i [2 \pi f_{\mathrm{car}}^M t+m_{\mathrm{sb}} \cdot E_{\mathrm{sb}}^M(t)]} \\
& \approx \sqrt{P_M} \mathrm{J}_0\left(m_{\mathrm{sb}}\right) e^{i \left(2 \pi f_{\mathrm{car}}^M  t\right)} \\
& +i \sqrt{P_M} \mathrm{J}_1\left(m_{\mathrm{sb}}\right) e^{i \left[2 \pi f_{\mathrm{up}}^M  t+m_{\mathrm{prn}}\cdot\phi _{\mathrm{seq}}^{M}(t)\right]} \\
& +i \sqrt{P_M} \mathrm{J}_1\left(m_{\mathrm{sb}}\right) e^{i \left[2 \pi f_{\mathrm{low}}^M t-m_{\mathrm{prn}} \cdot \phi _{\mathrm{seq}}^{M}(t)\right]}
\end{split}
\end{equation}
where $P_{M}$ is the power of the master laser, $f_{\mathrm{car}}^M$ is the laser frequency, $f_{\mathrm{up}}^M = f_{\mathrm{car}}^M + f_{\mathrm{sb}}^M$ is the frequency of the upper sideband of the modulated laser signal and $f_{\mathrm{low}}^M = f_{\mathrm{car}}^M - f_{\mathrm{sb}}^M$ is the frequency of the lower sideband. The higher‑order expansion terms of the modulated laser signal are omitted in Eq.~\eqref{eq:2} due to their negligible amplitudes.
At the slave spacecraft, the incoming laser from the master spacecraft interferes with the local modulated laser signal $E_{\mathrm{car}}^S(t)$ with power $P_S$, then the resulting interferometric signal passing through the photodetector is given as following
\begin{equation}\label{eq:3}
\begin{split}
V(t) &= \frac{\eta G_{\mathrm{TIA}} \sqrt{\gamma}}{N_{\mathrm{PD}}} \cdot \mathrm{Re}\left(E_{\mathrm{car}}^M(t) \cdot E_{\mathrm{car}}^{S^*}(t)\right) \\
&\approx G_t  \mathrm{J}_0^2\left(m_{\mathrm{sb}}\right) \cdot \cos\left(2\pi f_{\mathrm{car}}^h t\right) \\
& + G_t  \mathrm{J}_1^2\left(m_{\mathrm{sb}}\right) \cdot \cos\left[2\pi f_{\mathrm{sb}}^{\mathrm{up}}  t+m_{\mathrm{prn}}\cdot\phi_{\mathrm{seq}}(t)\right] \\
& + G_t  \mathrm{J}_1^2\left(m_{\mathrm{sb}}\right) \cdot \cos\left[2\pi f_{\mathrm{sb}}^{\mathrm{low}}  t-m_{\mathrm{prn}}\cdot\phi_{\mathrm{seq}}(t)\right]
\end{split}
\end{equation}
where $f_{\mathrm{car}}^h$ and $f_{\mathrm{sb}}^h$ respectively denote the differential frequencies between the laser carriers and USO clock signals of the two spacecraft, and $\phi_{\mathrm{seq}}(t)=\sum\limits_{i=1}^{L} (C_i^M-C_i^S) \cdot p(t-i T)$ is the phase difference between the data‑encoded PRN signals of the two spacecraft. Thus, the frequencies of the upper and lower sidebands in the interferometric signal are denoted as $f_{\mathrm{sb}}^{\mathrm{up}} = f_{\mathrm{car}}^h+f_{\mathrm{sb}}^h$ and $f_{\mathrm{sb}}^{\mathrm{low}} = f_{\mathrm{car}}^h-f_{\mathrm{sb}}^h$. Moreover, $G_t=\eta G_{\mathrm{TIA}} \sqrt{\gamma  P_MP_S}/ N_{\mathrm{PD}}$, with $\eta$ being the responsivity of the photodetector, $G_{\mathrm{TIA}}$ the transimpedance amplifier gain, $\gamma$ the heterodyne detection efficiency, and $N_{\mathrm{PD}}$ the number of segments in one photodiode~\cite{Yamamoto2023ClockSync, Esteban2012LaserPhD}.
The noise of the data‑encoded PRN signal on the system's laser phase measurement is the signal component at the main carrier frequency of the interferometric signal, equivalent to that of the baseband data‑encoded PRN signal at $f_{\mathrm{sb}}^h$. Due to the simultaneous appearance of the data‑encoded PRN signal on both sidebands and an amplitude ratio of $\mathrm{J}_0^2\left(m_{\mathrm{sb}}\right):\mathrm{J}_1^2\left(m_{\mathrm{sb}}\right)$ between the main carrier and single sideband, the laser phase measurement noise induced by the data‑encoded PRN signal in the proposed integrated scheme can be calculated according to Eq.~\eqref{eq:3} as
\begin{equation}\label{eq:4}
\begin{split}
\Phi_{\mathrm{sb}}^{\mathrm{car}}(f)=\sqrt{2} \cdot\frac{\mathrm{J}_1^2\left(m_{\mathrm{sb}}\right)}{\mathrm{J}_0^2\left(m_{\mathrm{sb}}\right)} m_{\mathrm{prn}} \cdot \phi_{\mathrm{prn}}({f+f_{\mathrm{sb}}^h})
\end{split}
\end{equation}
where $\phi_{\mathrm{prn}}(f)$ refers to the amplitude spectral density (ASD) of the unit‑amplitude data‑encoded PRN signal at frequency $f$. For the currently‑adopted carrier phase modulation based integrated scheme~\cite{Yamamoto2024LISA_ranging, Heinzel2011, Sutton2013OpticalRanging}, the laser phase measurement noise from the data‑encoded PRN signal is expressed as
\begin{equation}\label{eq:5}
\begin{split}
\Phi_{\mathrm{car}}^{\mathrm{car}}(f)=m_{\mathrm{prn}} \cdot \phi_{\mathrm{prn}}(f)
\end{split}
\end{equation}
For $f$ in the frequency band of interest for space‑based GW detection, i.e., from $0.1\,\mathrm{mHz}$ to $1\,\mathrm{Hz}$, $\phi_{\mathrm{prn}}(f)$ is significantly reduced when frequency‑shifted by $f_{\mathrm{sb}}^h$, such that $\phi_{\mathrm{prn}}({f+f_{\mathrm{sb}}^h}) \ll \phi_{\mathrm{prn}}(f)$. Furthermore, since $m_{\mathrm{sb}}$ satisfies $\frac{\mathrm{J}_1^2\left(m_{\mathrm{sb}}\right)}{\mathrm{J}_0^2\left(m_{\mathrm{sb}}\right)} < 1$, it follows that $\Phi_{\mathrm{sb}}^{\mathrm{car}}(f) \ll \Phi_{\mathrm{car}}^{\mathrm{car}}(f)$. Compared with the currently‑adopted carrier phase modulation based integrated scheme, the majority of the spectral energy of the data‑encoded PRN signal is shifted outside the laser phase measurement bandwidth by phase‑modulating the data‑encoded PRN signal onto the USO clock signal in the scheme proposed here. Overlap between the main carrier and the main lobe of the data‑encoded PRN signal is avoided in this scheme (as shown in Fig.~\ref{fig1}), and thus the noise on laser phase measurement induced by the signal is suppressed.
Clock noise transfer is implemented by measuring the phases of the clock sidebands within the interferometric signal, which are contaminated by the noise component of the data‑encoded PRN signal at the sideband frequencies. According to the signal model in Eq.~\eqref{eq:3}, the expression for the sideband phase measurement noise induced by the data‑encoded PRN signal can be given as
\begin{equation}\label{eq:6}
\begin{split}
\Phi_{\mathrm{sb}}^{\mathrm{sb}}(f)=m_{\mathrm{prn}} \cdot \phi_{\mathrm{prn}}(f)
\end{split}
\end{equation}
For clock noise transfer of the currently‑adopted carrier phase modulation based integrated scheme, the noise from the data‑encoded PRN signal is expressed as
\begin{equation}\label{eq:7}
\begin{split}
\Phi_{\mathrm{car}}^{\mathrm{sb}}(f)=\frac{\mathrm{J}_0^2\left(m_{\mathrm{sb}}\right)}{\mathrm{J}_1^2\left(m_{\mathrm{sb}}\right)} m_{\mathrm{prn}} \cdot \phi_{\mathrm{prn}}({f+f_{\mathrm{sb}}^h})
\end{split}
\end{equation}
Evidently, clock noise transfer of the system experiences greater noise from the data‑encoded PRN signal in the proposed integrated scheme. To address this problem, the modulation depths should be appropriately designed to confine the noise to meet system requirements. Moreover, as shown in Fig.~\ref{fig5}, employing optimized PRN codes such as BBC for the proposed integrated scheme can also effectively reduce this impact.
Additionally, in the signal processing architecture of a phase‑locked loop (PLL) cascaded with a delay‑locked loop (DLL), the data‑encoded PRN signal loses its low‑frequency components during phase measurement due to the high‑pass filtering property of the PLL~\cite{Sutton2010, Euringer2024}. However, as the bandwidth of the sideband PLL in the detection system is much narrower than that of the carrier PLL~\cite{2023OExpr..3134648Z}, the data‑encoded PRN signal fed into the DLL in the proposed integrated scheme retains more low‑frequency components, minimizing PLL's impact on the system’s ranging and communication performance in comparison with the currently‑adopted integrated scheme.
\section{Simulation Results}
\begin{figure}[ht!]
 \centering
        \includegraphics[width=8.5cm,keepaspectratio]{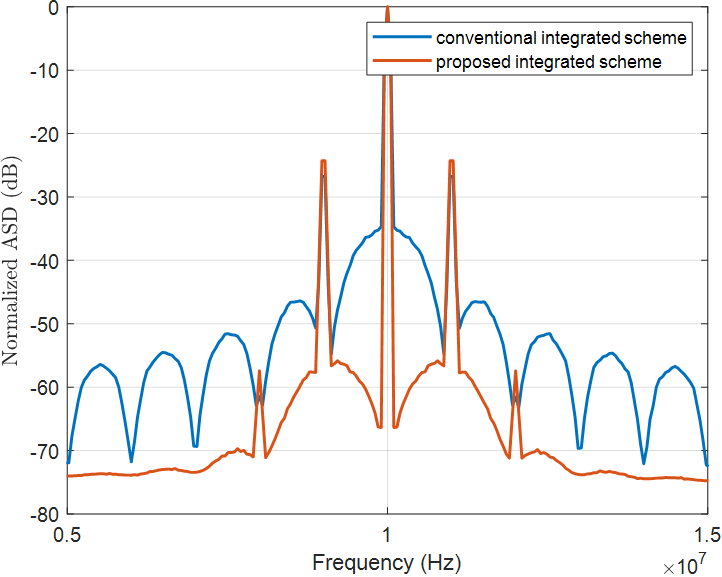}
 \caption{Normalized ASD of the interferometric signals for the currently‑adopted carrier phase modulation based integrated scheme and the sideband modulation based scheme of inter‑satellite integrated laser link.}
\label{fig3}
\end{figure}

\begin{figure}[ht!]
 \centering
        \includegraphics[width=8.5cm,keepaspectratio]{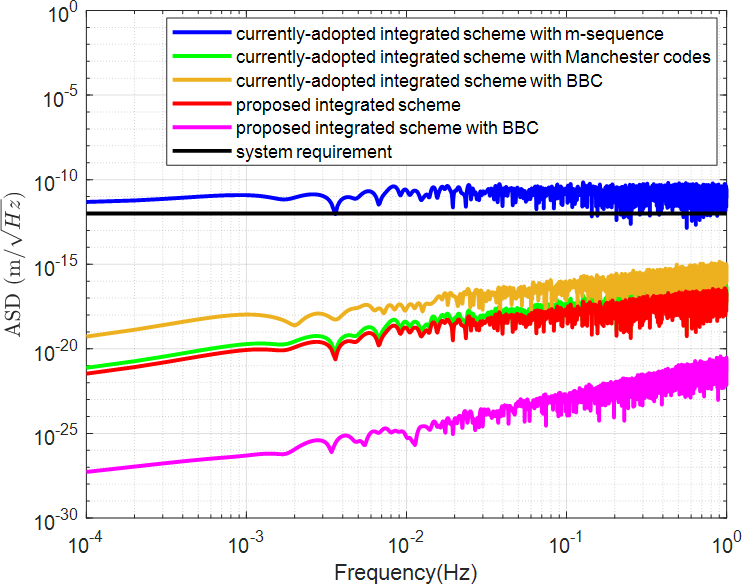}
 \caption{ASD of the laser phase measurement noise induced by the data‑encoded PRN signal for the two schemes with three PRN code types.}
\label{fig4}
\end{figure}

\begin{figure}[ht!]
 \centering
        \includegraphics[width=8.5cm,keepaspectratio]{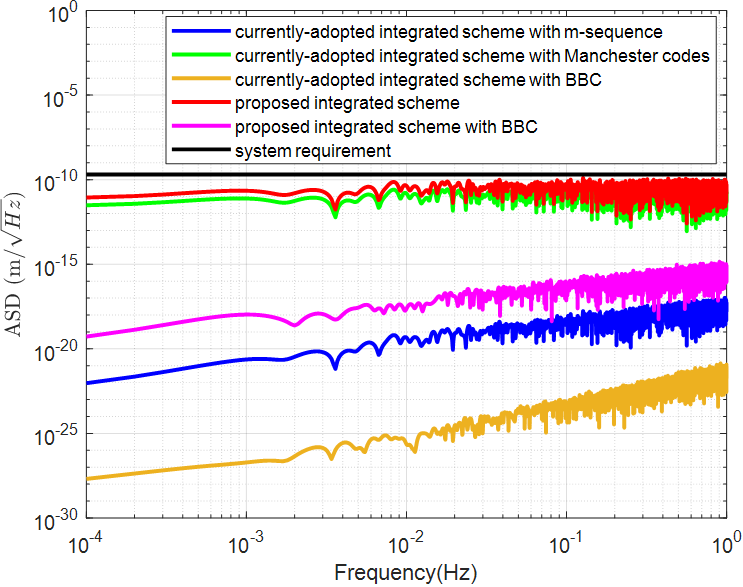}
 \caption{ASD of the sideband phase measurement noise caused by the data‑encoded PRN signal for different schemes with three PRN code types.}
\label{fig5}
\end{figure}

To support the theoretical analysis of the proposed scheme of inter‑satellite integrated laser link, the simulations of ASD of the noise induced by the data‑encoded PRN signal on laser phase measurement and clock noise transfer of the system are carried out here with the parameters selected as shown in Table~\ref{table1}. The PRN chip rate is set to match the frequency difference of the USO clock signals from the two spacecraft so that the main carrier of the interferometric signal is shifted to the exact frequency where the data‑encoded PRN signal has its minimum ASD. This minimizes the data‑encoded PRN signal‑induced noise on laser phase measurement, i.e., $\phi_{\mathrm{prn}}({f+f_{\mathrm{sb}}^h}) \approx 0$, for $f$ from $0.1\,\mathrm{mHz}$ to $1\,\mathrm{Hz}$. On the other hand, the modulation depths are designed such that the signal components in the modulated laser signal transmitted by the spacecraft maintain a power distribution ratio of 100:15:1 among the laser carrier, clock sidebands, and the data‑encoded PRN signal, consistent with that of the currently‑adopted integrated scheme. From a noise perspective, laser frequency noise, a form of phase noise, is the dominant noise component within the interferometric signal. As a result, the signal fed into the DLL achieves higher signal‑to‑noise ratio (SNR) under the proposed integrated scheme with larger PRN modulation depth.
\begin{table}
\caption{Parameters of the proposed integrated scheme\label{table1}}
\centering
\renewcommand{\arraystretch}{1.3}
\begin{tabular}{ccc}
\hline
Parameter & Value & Comment \\
\hline
\begin{tabular}[c]{@{}c@{}}Laser frequency on the \\ remote SC $f_{\mathrm{car}}^M$\end{tabular} & 281.95489\,THz & \multirow{2}{*}{\begin{tabular}[c]{@{}c@{}}Carrier beatnote's \\ frequency is 10 \\ MHz\end{tabular}} \\
\begin{tabular}[c]{@{}c@{}}Laser frequency on the \\ local SC $f_{\mathrm{car}}^S$\end{tabular} & 281.95488\,THz & \\
\hline
\begin{tabular}[c]{@{}c@{}}USO clock frequency \\ on the remote SC $f_{\mathrm{sb}}^M$\end{tabular} & 2.001\,GHz & \multirow{2}{*}{\begin{tabular}[c]{@{}c@{}}Sideband \\ beatnotes are 9 \\ MHz and 11\,MHz\end{tabular}} \\
\begin{tabular}[c]{@{}c@{}}USO clock frequency \\ on the local SC $f_{\mathrm{sb}}^S$\end{tabular} & 2.000\,GHz & \\
\hline
System clock rate $f_{s}$ & 100\,MHz & \\
\hline
\begin{tabular}[c]{@{}c@{}}PRN modulation depth \\ $m_{\mathrm{prn}}$\end{tabular} & 0.258\,rad & \multirow{2}{*}{\begin{tabular}[c]{@{}c@{}}The power \\ distribution \\ is 100:15:1\end{tabular}} \\
\begin{tabular}[c]{@{}c@{}}Clock‑sideband modulation \\ depth $m_{\mathrm{sb}}$\end{tabular} & 0.528\,rad & \\
\hline
PRN code length $L$ & 1024 & \\
PRN chip rate $f_{\mathrm{chip}}$ & 1\,MHz & \\
Raw data rate $f_{\mathrm{bit}}$ & 62.5\,kbps & \\
Spreading factor $SF$ & 16 & \\
\hline
\end{tabular}
\end{table}
In Fig.~\ref{fig3}, normalized ASD curves of the interferometric signals are plotted based on the signal model and parameters for the two integrated schemes. Since the power of the data‑encoded PRN signal is concentrated around the main carrier in the currently‑adopted integrated scheme, the accuracy of laser phase measurement is significantly degraded. In contrast, for the proposed integrated scheme, the power of the data‑encoded PRN signal is mostly distributed around the clock sidebands on both sides apart from the main carrier, with its power leakage at the main carrier frequency reaching a minimum. As a result, the noise on laser phase measurement from the data‑encoded PRN signal in the proposed integrated scheme is greatly reduced.
Fig.~\ref{fig4} presents the level of noise induced by the data‑encoded PRN signal on laser phase measurement, which is quantitatively evaluated for the two schemes with three distinct PRN code types, namely the m‑sequence currently in common use~\cite{Xie_2023}, Manchester codes, and BBC. The evaluation is performed according to Eqs.~\eqref{eq:4} and \eqref{eq:5}, with the results displayed. As shown in the figure, for the currently‑adopted integrated scheme with m‑sequence, the laser phase measurement noise induced by the data‑encoded PRN signal exceeds the $1\,\mathrm{pm/\sqrt{Hz}}$ requirement of the detection system (the black line) by nearly one order of magnitude. For comparison, when the data‑encoded PRN signal with the identical PRN sequence is phase‑modulated onto the USO clock signal, the laser phase measurement noise is substantially suppressed to meet the system requirement. Specifically, the noise level is reduced by nearly five orders of magnitude, yielding a lower noise level than the currently‑adopted integrated scheme employing the two improved code types. Notably, the noise reaches its minimum when BBC is adopted in the proposed integrated scheme (the magenta curve), which confirms that the millihertz‑band component of the data‑encoded PRN signal no longer constitutes a limiting factor for laser phase measurement performance of the system.
As illustrated in Fig.~\ref{fig5}, the magnitude of noise from the data‑encoded PRN signal on clock noise transfer is quantitatively evaluated for the two schemes with the three PRN code types using Eqs.~\eqref{eq:6} and \eqref{eq:7}, and the results are plotted. It can be observed that the sideband phase measurement noise induced by the data‑encoded PRN signal is higher in the proposed integrated scheme using m‑sequence than that in the currently‑adopted one, yet it remains safely below the $ f_{\mathrm{sb}}/f_{\mathrm{car}}^h \times 1\,\mathrm{pm/\sqrt{Hz}}$ requirement (the black line), and thus imposes no limitation on the clock noise transfer accuracy of the system~\cite{Barke:2010}. Furthermore, the noise of the data‑encoded PRN signal is significantly reduced when the proposed integrated scheme with BBC is adopted, demonstrating excellent noise suppression performance.
As the discussion in this Letter, to reduce the noise from the data‑encoded PRN signal for ranging and communication, a sideband modulation based scheme of inter‑satellite integrated laser link for space‑based GW detection is proposed, in which the data‑encoded PRN signal is phase‑modulated onto the USO clock signal prior to its modulation onto the laser carrier. Under the proposed integrated scheme, the signal‑induced noise on the main carrier phase measurement is suppressed well below the system requirement and no longer limits the laser phase measurement performance of the system. This scheme provides a high‑performance integrated solution compatible with space‑based GW detectors, effectively eliminating the constraint imposed by data‑encoded PRN modulation on the system’s phase measurement accuracy. For practical implementation, reasonable modulation depths for the PRN signal and USO clock signal should be properly set to balance phase noise suppression and demodulation quality. Additionally, practical PRN code types require careful selection through simulations and experimental validation with the scheme, which is under study.
\ack{The authors would like to thank Professors C. Shao and Z. Yi for beneficial discussions. This work was fully supported by the National Key R\&D Program of China under Grant No.~2023YFC2205500, Research on Inter‑satellite Laser Ranging and Communication Technology.}

\end{document}